# Piezoelectric-Metal Phononic Crystal Enabling GHz Tunable Ultrahigh Q Quasi-BIC mode

Xuankai Xu[1,†], Jiawei Li[1,†], Ruoyu Wang[2,†], Ruihong Xiong[1], Yiwei Wang[1], Xiaoqin Shen[2,*], Tao Wu[1,3,*]

[1]School of Information Science and Technology, ShanghaiTech University, Shanghai, 201210, China
[2]School of Physical Science and Technology, ShanghaiTech University, Shanghai, 201210, China
[3]Shanghai Engineering Research Center of Energy Efficient and Custom AI IC, Shanghai, 201210, China

[†]*These authors contributed equally*
*Correspondence to: shenxq@shanghaitech.edu.cn; wutao@shanghaitech.edu.cn;

## Abstract

The integration of GHz-frequency, high quality factor ($Q$), and electrically tunable acoustic resonators holds significant potential for advancing applications in quantum information technologies, microwave photonics, and reconfigurable RF systems. However, simultaneously achieving these three characteristics within a single, scalable platform remains a fundamental challenge. Here, we report the experimental demonstration of a GHz quasi-BIC resonator in a piezoelectric thin-film shear horizontal (SH) wave system, achieved through a structurally simple piezoelectric–metal phononic crystal (PnC) architecture on a LiNbO$_3$ thin film. This approach enables leaky Fabry–Perot coupling mode and localized trapping quasi-BIC mode. Without the need for deep etching or intricate patterning, we achieve a room-temperature quality factor of $6\times10^4$ at ~1 GHz in ambient air, corresponding to an $f \times Q$ product of $6 \times 10^{13}$ Hz at quasi-BIC mode. Furthermore, we demonstrate efficient electrical tunability via low-voltage (0.6 V) electrothermal modulation of the PnC structure, enabling a reversible transition between trapped and transmission states and yielding a high-contrast amplitude modulation of 47.75 dB. Our results establish a lithography-friendly, fabrication-tolerant platform for realizing tunable, high-$Q$ acoustic resonators at GHz frequencies, overcoming longstanding barriers in phononic device engineering. This work opens new directions for scalable on-chip phononic circuits in quantum acoustics, reconfigurable RF systems, and signal processing applications.

## Introduction

In the rapidly evolving fields of quantum information processing, microwave photonics, and advanced RF systems, acoustic resonators that simultaneously achieve GHz operation, ultrahigh quality factors ($Q$), and electrical tunability are highly desirable. GHz frequencies are crucial for superconducting qubits[1–3], RF communications[4–6], and high-speed sensing[7–9], while high-$Q$ resonators enhance spectral selectivity[10,11] and signal-to-noise ratio[12,13]. Electrical tunability enables dynamic control in reconfigurable filters[14,15], modulators[16,17], and switches[18,19]. The integration of these three key attributes in a single device could unlock transformative capabilities in scalable quantum transducers, reconfigurable RF front-ends, and chip-scale phononic processors, offering precise control over acoustic wave propagation and real-time adaptability[20–22]. Yet, devices that combine all three remain rare, posing a major obstacle to the advancement of integrated phononic platforms in both classical and quantum technologies.

Bound states in the continuum (BICs) offer a promising route to overcome $Q$ factor limitations by suppressing radiative losses through symmetry protection or destructive interference, theoretically allowing infinite-$Q$ modes[23–27]. Despite their success in optics[28–30] and bulk acoustic systems[31,32], realizing BICs in the microwave-frequency regime of piezoelectric thin-film devices remains highly challenging. Most existing implementations rely on complex etched structures, such as partially etched phononic crystal (PnC) slabs or fully etched PnC holes/pillars[33–36], which require high fabrication precision. In practice, pattern misalignments and etching sidewall angles introduce scattering losses[33,37–39], severely constraining achievable $Q$ factors. Moreover, these complex structures pose significant fabrication challenges at higher frequencies, making it difficult to scale BIC resonators into the multi-GHz range. The static nature of such etched architectures also hinders real-time electrical control, limiting their applicability in dynamic RF and quantum systems. Thus, there is a critical need for a structurally simple, fabrication-friendly, and electrically tunable BIC platform that can simultaneously achieve ultrahigh $Q$ and GHz operation in practical, chip-scale devices.

In this work, we report the first experimental realization of a GHz quasi-BIC resonator in a piezoelectric thin-film shear horizontal (SH) wave system, leveraging a simple piezoelectric-metal PnC architecture. By patterning a metal-PnC lattice atop a LiNbO$_3$ thin film, we engineer phononic band structures that support both leaky Fabry–Perot coupling modes and localized trapping states. Through interference between the fundamental and higher-order SH modes, we achieve quasi-BIC formation by decoupling the acoustic energy from the radiation channel. Without resorting to partial etching or intricate structural tuning, we achieve an exceptionally high $Q$ factor of $6 \times 10^4$ at ~ 1 GHz in ambient air, yielding an $f \times Q$ product of $6 \times 10^{13}$ Hz — comparable to or surpassing the best reported values in far more complex systems. More importantly, we demonstrate that the tunability of quasi-BIC state. By applying a low-voltage (0.6 V) electrothermal bias to the piezoelectric-metal PnC, we modulate the acoustic interference condition and reversibly transition the mode from trapped to transmission, which results in a high-contrast amplitude modulation of 47.75 dB. Our work proves that ultrahigh-$Q$ and electrically tunable quasi-BIC modes can be realized using a structurally simple, lithography-friendly design, bypassing many of the limitations that have stalled prior efforts. This approach offers a practical, scalable platform for high-$Q$ filters, reconfigurable phononic networks, and quantum signal processors, bringing BIC-enabled acoustics closer to real-world integration.

## Results

**Design and phononic confinement of the metal–piezoelectric PnC resonator.** To realize GHz-frequency acoustic resonances energy coupling channel, we designed a one-

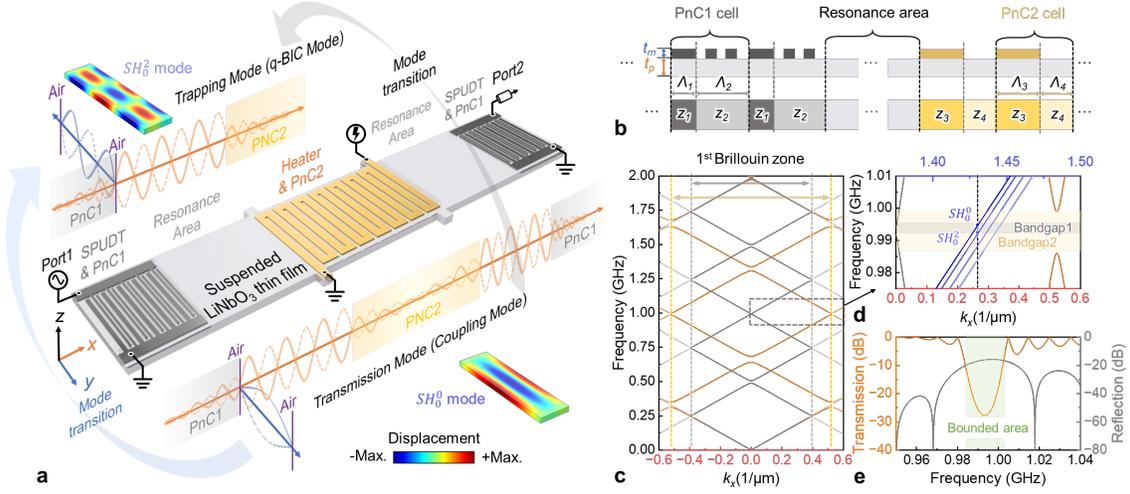

**Fig. 1 Design of the metal–piezoelectric PnC resonator. a** Schematic of the one-dimensional (1D) metal–piezoelectric PnC resonator. SH waves are generated and detected by SPUDTs via the piezoelectric effect. The SPUDTs also serve as outer reflectors (PnC1), forming a phononic bandgap. The inner reflectors (PnC2) define the cavity boundary and simultaneously function as integrated heaters for thermoelastic modulation. **b** Cross-sectional view of the resonant cavity, showing the periodic arrangement of PnC1 and PnC2. **c** Calculated band structures of PnC1 (grey) and PnC2 (orange), exhibiting overlapping bandgaps centered near 1 GHz. This spectral overlap enables confinement of acoustic waves within the cavity. **d** Selective trapping of shear horizontal standing wave modes, including $SH_0^0$ (no nodal points along the y-direction) and $SH_0^2$ (two nodal points along the y-direction) mode, within the overlapping bandgap region. **e** Simulated reflection spectrum of PnC1 (10 unit cells) and transmission spectrum of PnC2 (31 unit cells). The shaded region indicates the overlapping bandgap that defines the acoustic confinement window.

dimensional (1D) metal–piezoelectric phononic crystal (PnC) resonator based on a suspended Z-cut LiNbO₃ thin film. As illustrated in Fig. 1a, shear horizontal (SH) waves are excited and detected by single-phase unidirectional transducers (SPUDTs), which also serve as outer reflectors (PnC1), forming a phononic bandgap. The inner reflectors (PnC2) define the cavity boundaries and simultaneously act as electrothermal heaters, enabling dynamic tuning of the acoustic response (Fig. 1b). Detailed design parameters are provided in Supplementary Note 1. Fig. 1c shows the simulated band structures of PnC1 (grey) and PnC2 (orange), both exhibiting phononic bandgaps centered near 1 GHz. Their spectral overlap enables strong confinement of SH waves within the central cavity. In Fig. 1d, the blue lines represent lateral SH mode dispersion in a 70 μm wide, 290 nm thick LiNbO₃ plate, while the black dashed line marks the in-plane wavevector $k_x$ selected for resonance. Two confined lateral modes — fundamental $SH_0^0$ and higher-order $SH_2^0$ — reside within the bandgap and are distinguished by their number of lateral nodal lines. The overlapping bandgap enables the formation of two distinct resonant behaviors. When the $SH_0^0$ mode couples between the two resonators through PnC2, it forms a transmission mode that leaks into the surrounding continuum. However, when the $SH_0^0$ and $SH_2^0$ modes interfere destructively, their hybridization suppresses radiation leakage, forming a confined quasi-

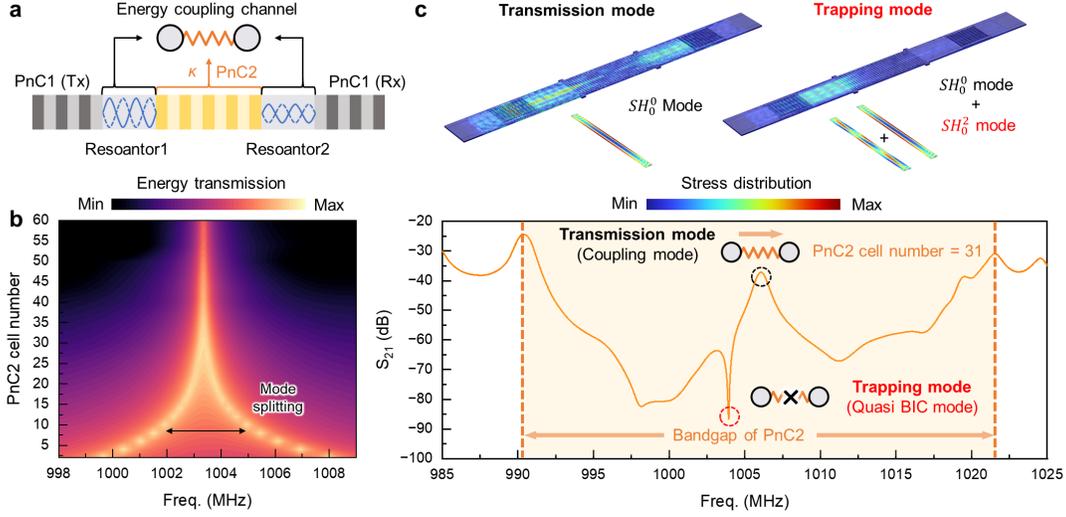

**Fig. 2 Formation of a quasi-BIC via engineered phononic coupling. a** Two SH wave resonators (Resonator 1 and Resonator 2) are coupled through a central PnC2 lattice. PnC1 (left and right) defines the resonator boundaries by reflecting SH waves, while PnC2 serves as a tunable acoustic spring that mediates inter-cavity coupling. **b** A reduced number of unit cells weakens confinement and increases the inter-resonator coupling strength (κ), resulting in mode splitting. **c** Simulated frequency response within the PnC2 bandgap (shaded region) demonstrates a transmission channel that supports coupling of the $SH_0^0$ mode between the two resonators. Destructive interference between the $SH_0^0$ and $SH_0^2$ modes suppresses radiation loss, decoupling from the radiation channel, enabling the formation of a hybridized quasi-BIC with an enhanced $Q$. Simulated stress profiles (top) illustrate the transmission mode with delocalized energy (left) and the quasi-BIC mode with localized energy confinement (right).

BIC. As in Friedrich–Wintgen BICs, this destructive interference results in an avoided crossing of the coupled modes and the emergence of a high-$Q$ resonance trapped inside the bandgap[40]. Fig.1e confirms the bandgap overlap and the selective trapping behavior through simulated reflection and transmission spectra.

**Engineering quasi-BIC mode through phononic coupling.** To construct interference-induced quasi-BIC modes, we implemented a coupled-resonator structure consisting of two SH wave cavities connected via a central PnC2 region (Fig. 2a). In this configuration, the outer PnC1 segments function as reflective boundaries, while the intermediate PnC2 serves as a phononic coupler. By tuning the number of PnC2 unit cells, the effective stiffness of the coupling region can be adjusted, allowing the construction of a coupled Fabry-Pérot resonator. The detailed design of this configuration is provided in Supplementary Note 2. As the number of PnC2 unit cells is reduced (Fig. 2b), the inter-cavity coupling strength $\kappa$ increases, enhancing modal interaction and leading to mode splitting. Finite-element simulations (Fig. 2c) show that, within the overlapping phononic bandgap, the fundamental $SH_0^0$ mode supports a transmission channel across the coupled cavities. However, when the $SH_0^0$ and higher-order $SH_2^0$ modes become nearly degenerate, their destructive interference gives rise to a hybridized mode that decouples from the continuum — forming a quasi-BIC with a substantially enhanced $Q$ of the resonator. The simulated stress field distributions illustrate the distinction between a delocalized transmission mode and a spatially confined quasi-BIC mode, confirming the role of phononic coupling and modal interference in achieving acoustic confinement.

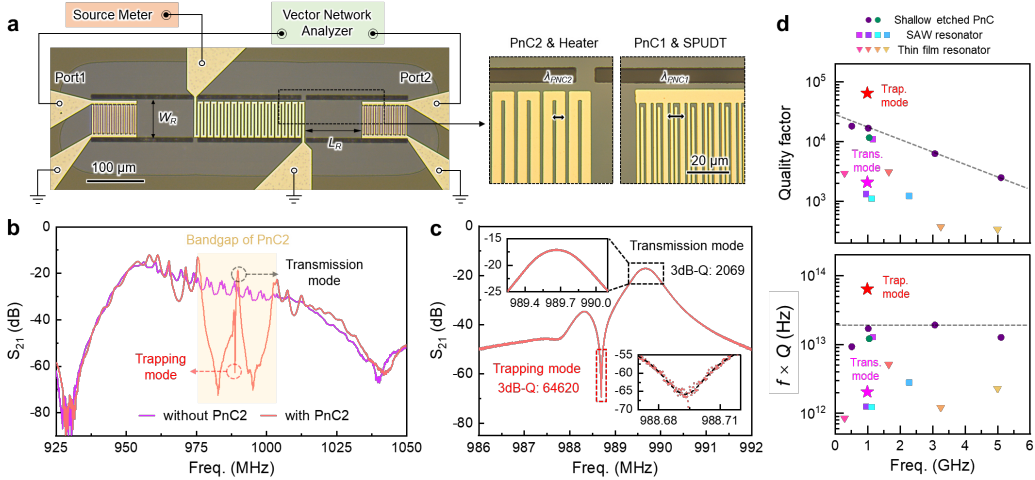

**Fig. 3 Room-temperature transmission spectroscopy of the metal–piezoelectric PnC resonator. a** Optical micrograph of the fabricated resonator, with inset images highlighting the periodic structures of PnC1 and PnC2. Transmission measurements ($S_{21}$) are performed using a vector network analyzer (VNA), while a source meter supplies heating power to enable thermal tuning of the PnC2 lattice. **b** Measured $S_{21}$ spectra with and without the PnC2 structure (0 V input from the source meter). **c** Magnified view of the $S_{21}$ spectra showing a substantial $Q$-factor enhancement—from 2,069 (transmission mode) to 64,620 (quasi-BIC mode). **d** Comparison of the achieved $Q$-factor (highlighted by a star) with state-of-the-art $LiNbO_3$-based resonators reported in the literature, and benchmarking of the frequency–quality factor product ($f \times Q$) for the same set of devices.

**Experimental observation of high-Q acoustic trapping.** We experimentally validated the quasi-BIC resonator design through room-temperature transmission measurements on a fabricated device (Fig. 3a). The resonance region has a length $L_R = 101.5$ μm and a width = 70 μm; the unit cell lengths of PnC1 and PnC2 are $\lambda_{PnC1} = 8$ μm and $\lambda_{PnC2} = 6$ μm, respectively. Transmission spectra ($S_{21}$) were recorded using a vector network analyzer (VNA), while a source meter supplied DC heating power to the integrated PnC2 heaters. Further details on device fabrication and the experimental setup are provided in the Methods section. As shown in Fig. 3b, the control device without PnC2 exhibits no bandgap, allowing free propagation of SH waves. In contrast, incorporation of PnC2 leads to the formation of a clear transmission bandgap (shaded region), within which two distinct resonances emerge: a broad transmission mode (coupling mode) and a narrow trapping mode (quasi-BIC mode). Comparative measurements with and without PnC2 are presented in Supplementary Note 2. A magnified view of $S_{21}$ (Fig. 3c) reveals a dramatic quality factor enhancement—from 2,069 for the transmission mode to 64,620 for the quasi-BIC mode—providing direct evidence of radiation suppression via engineered modal interference. Benchmarking against state-of-the-art $LiNbO_3$-based resonators[35,36,41–48] (Fig. 3d) shows that our device achieves an $f \times Q$ product exceeding $6 \times 10^{13}$ Hz, comparable to the best reported values, while maintaining a fabrication-friendly architecture that avoids deep etching or subwavelength patterning.

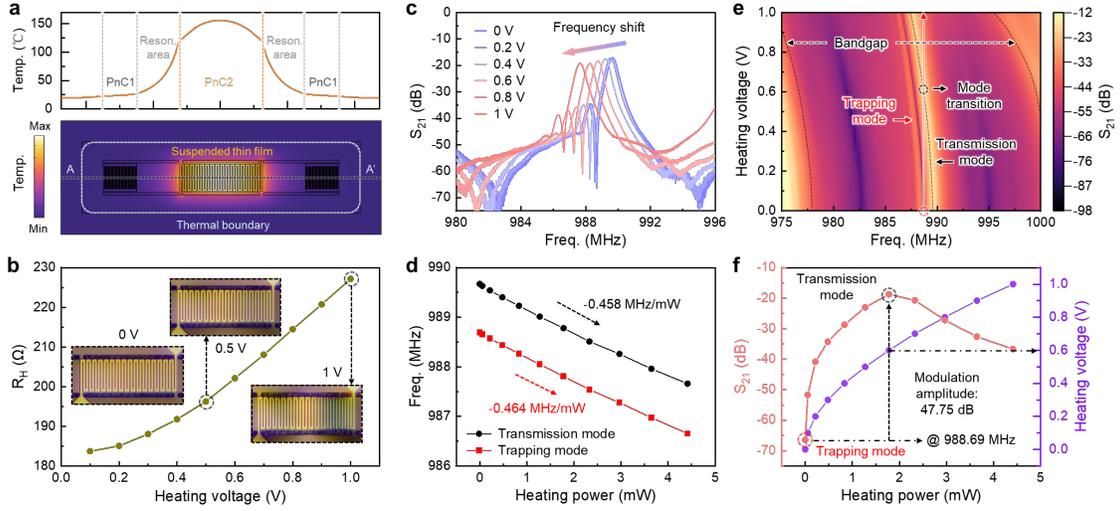

**Fig. 4 Thermal tuning and dynamic modulation. a** Simulated thermal profile of the resonator under Joule heating, illustrating the localized temperature distribution: PnC2 acts as the central heating zone, the resonance region is moderately heated, while PnC1 remains near ambient temperature. **b** Measured resistance of the integrated heater as a function of input voltage. Optical micrographs show progressive thermal expansion of the PnC2 region with increasing voltage (0 V to 1 V). **c** Transmission mode, and quasi-BIC mode redshift as the applied heating voltage increases. **d** Extracted thermal sensitivity of the quasi-BIC and transmission modes, showing frequency shifts of -0.464 MHz/mW and -0.458 MHz/mW, respectively. **e** Modulation spectra under stepwise voltage application demonstrate redshifts of both the PnC2 bandgap and resonance modes. **f** Mode transition under thermal tuning, the quasi-BIC transitions into the transmission mode at 0.6 V, at 988.69 MHz.

**Dynamic tuning of the transmission and trapping modes.** To evaluate the active tunability of the resonator, we applied voltage-controlled Joule heating to the PnC2 region and monitored the resulting thermal and spectral response. Simulated thermal profiles (Fig. 4a) reveal that heating is localized within PnC2, with the resonance cavity moderately heated and the PnC1 reflectors remaining near ambient temperature. The resistance of the integrated heater increases with voltage (Fig. 4b), while optical micrographs show progressive thermal expansion of PnC2 from 0 V to 1 V. Thermal tuning results in consistent frequency shifts of both the transmission and trapping modes (Fig. 4c), with extracted thermal sensitivities of -0.458 MHz/mW and -0.464 MHz/mW, respectively (Fig. 4d). As shown in Fig. 4e, stepwise voltage application shifts both the phononic bandgap and the resonance frequencies, with the bandgap exhibiting the largest shift due to direct heating. Most notably, thermal modulation induces a transition from the trapping mode to a transmission mode at an input voltage of 0.6 V (Fig. 4f), accompanied by a dramatic change in transmission amplitude. At 988.69 MHz, the device achieves a modulation depth of 47.75 dB, demonstrating its potential for low-power signal control in reconfigurable RF and phononic systems.

## Discussion

In this work, we demonstrate a structurally simple yet highly effective approach to achieving GHz-frequency, ultrahigh-$Q$, and electrically tunable acoustic resonators using a metal-piezoelectric PnC on a suspended LiNbO$_3$ thin film. Through phononic bandgap engineering and modal interference between SH modes, we realize a quasi-BIC resonance with a $Q$-factor exceeding $6 \times 10^4$ and an $f \times Q$ product exceeding $6 \times 10^{13}$ Hz in room temperature — comparable to or better than more complex etched systems. Meanwhile, the PnC2 region also functions as an integrated electrothermal modulator, enabling low-voltage, reversible tuning between trapped and transmission states. The observed 47.75 dB modulation depth and MHz/mW-level thermal sensitivities confirm the resonator potential for active control in reconfigurable RF and phononic circuits.

Unlike prior BIC-based devices requiring precise etching or nanofabrication, our design is lithography-friendly and fabrication-tolerant, making it scalable for large-area integration. The modularity of the architecture also supports future extensions to quantum acoustics, nonreciprocal devices, and programmable phononic networks. Overall, this work establishes a new platform for high-performance and tunable acoustic devices, combining simplicity, performance, and control in a form readily adaptable to practical applications.

## Methods

**Device Fabrication.** A 750 nm Z-cut LiNbO$_3$ thin film, bonded to a 4-inch silicon substrate and supplied by NGK Insulators, Ltd., was thinned down to 290 nm using ion beam etching (IBE). Device release boundaries were then defined and etched by IBE. Subsequently, a 100 nm aluminum top electrode was deposited via physical vapor deposition (PVD) and patterned using a lift-off process. Finally, the underlying silicon was selectively removed using XeF$_2$, forming a 60 μm-deep air cavity beneath the suspended LiNbO$_3$ membrane.

**Finite-Element Modeling**. We simulated the frequency response and stress distribution of transmission mode and coupling mode using finite-element methods with COMSOL Multiphysics software, a widely used tool for simulating acoustic thin-film resonators. The solid mechanics and electrostatic modules were coupled in the Piezoelectric Effect Multiphysics module, enabling computation of both mechanical and electrical responses. The software solves elastic wave equations in a linear piezoelectric medium with strain-charge coupling. Material properties for the simulation were sourced from the COMSOL Material Library.

**Experimental Setup**. The S-parameters of the devices were measured using a vector network analyzer (Keysight N5234B) with an RF probe station (MPI TS150) equipped with two GSG RF probes (MPI T26A, 150 μm pitch). For thermal tuning, a DC bias voltage was applied through two DC probes connected to a precision source measure unit (Keysight B2901A).

## Acknowledgements

This work was supported by the National Key Research and Development Program of China (Grant No. 2024YFB3614100), National Natural Science Foundation of China (Grant No. 62275152), Shanghai Natural Science Foundation (Grant No. 23ZR1442400), Jiangsu Provincial Key Research and Development program (Grant No. BE2023048), and Foundation of the Key State Laboratory of Transducer Technology (Grant No. SKT2303). The authors also appreciate the device fabrication and simulation support from ShanghaiTech Material Device Laboratory (SMDL) and the HPC Platform.

## Author Contribution

X. Xu conceived the concept, carried out theoretical analysis, contributed to device fabrication and measurements, and wrote the manuscript. J. Li. performed the finite element simulations and participated in device fabrication and measurements. R. Wang conducted phononic band structure calculations, contributed to theoretical discussions, and assisted in manuscript preparation. R. Xiong and Y. Wang assisted in device fabrication and measurements. X. Shen and T. Wu supervised the project, provided technical guidance, and revised the manuscript. All authors discussed the results and contributed to the final version of the manuscript. X. Xu, J. Li and R. Wang contributed equally to this work.

The authors declare no competing interests or conflicts of interest. Correspondence and requests for materials should be addressed to Xiaoqin Shen (shenxq@shanghaitech.edu.cn) and Tao Wu (wutao@shanghaitech.edu.cn).